\documentclass[%
preprint,
superscriptaddress,
showpacs,preprintnumbers,
nobibnotes,
amsmath,amssymb,
aps,
prl,
endfloats
]{revtex4-1}
\usepackage{textcomp}
\usepackage{graphicx}
\usepackage{dcolumn}
\usepackage{bm}

\begin{document}
	
	
	\title{Linear magnetoresistance in a quasi-free two dimensional electron gas in an ultra-high mobility GaAs quantum well }
\author{T.~Khouri}
\email{T.Khouri@science.ru.nl}
\affiliation{Radboud University, High Field Magnet Laboratory (HFML-EMFL), Toernooiveld 7, 6525 ED Nijmegen, The Netherlands} 
\affiliation{Radboud University, Institute of Molecules and Materials, Heyendaalseweg 135, 6525 AJ Nijmegen, Netherlands} 
\author{U.~Zeitler}
\affiliation{Radboud University, High Field Magnet Laboratory (HFML-EMFL), Toernooiveld 7, 6525 ED Nijmegen, The Netherlands} 
\affiliation{Radboud University, Institute of Molecules and Materials, Heyendaalseweg 135, 6525 AJ Nijmegen, Netherlands} 
\author{C.~Reichl}
\affiliation{Laboratory for Solid State Physics, ETH Z\"urich, 8093 Z\"urich, Switzerland}
\author{W.~Wegscheider}
\affiliation{Laboratory for Solid State Physics, ETH Z\"urich, 8093 Z\"urich, Switzerland}
\author{N.~E.~Hussey}
\affiliation{Radboud University, High Field Magnet Laboratory (HFML-EMFL), Toernooiveld 7, 6525 ED Nijmegen, The Netherlands} 
\affiliation{Radboud University, Institute of Molecules and Materials, Heyendaalseweg 135, 6525 AJ Nijmegen, Netherlands} 
\author{S.~Wiedmann}
\email{S.Wiedmann@science.ru.nl}
\affiliation{Radboud University, High Field Magnet Laboratory (HFML-EMFL), Toernooiveld 7, 6525 ED Nijmegen, The Netherlands} 
\affiliation{Radboud University, Institute of Molecules and Materials, Heyendaalseweg 135, 6525 AJ Nijmegen, Netherlands} 
\author{J.C.~Maan}
\affiliation{Radboud University, High Field Magnet Laboratory (HFML-EMFL), Toernooiveld 7, 6525 ED Nijmegen, The Netherlands} 
\affiliation{Radboud University, Institute of Molecules and Materials, Heyendaalseweg 135, 6525 AJ Nijmegen, Netherlands} 	
	\date{\today}
	
	\begin{abstract}
We report a magnetotransport study of an ultra-high mobility ($\bar{\mu}\approx 25\times 10^6$\,cm$^2$\,V$^{-1}$\,s$^{-1}$) $n$-type GaAs quantum well up to 33\,T. A strong linear magnetoresistance (LMR) of the order of 10$^5$\% is observed in a wide temperature range between 0.3\,K and 60\,K. The simplicity of our material system with a single sub-band occupation and free electron dispersion rules out most complicated mechanisms that could give rise to the observed LMR. 
At low temperature, quantum oscillations are superimposed onto the LMR. Both, the featureless LMR at high $T$ and the quantum oscillations at low $T$ follow the empirical resistance rule which states that the longitudinal conductance is directly related to the derivative of the transversal (Hall) conductance multiplied by the magnetic field and a constant factor $\alpha$ that remains unchanged over the entire temperature range. Only at low temperatures, small deviations from this resistance rule are observed beyond $\nu=1$ that likely originate from a different transport mechanism for the composite fermions.

	\end{abstract}
	
	\maketitle
	
	
Magnetoresistance studies are one of the simplest yet most powerful tools to probe the electronic properties of solids. For metals and semiconductors, classical theories predict the resistance to first increase quadratically with a transverse external magnetic field and then saturate \cite{Rossiter2006}. In a growing number of novel material systems including topological insulators (TIs) \cite{Wang2012,Gusev2013b,Wang2012a,Tang2011,Zhao2013_2,Wang2011_2}, Dirac \cite{Novak2015,Liang2014,Feng2015,Zhao2015} and Weyl semi-metals \cite{Shekhar2015,Shekhar,Zhao2015_2} and silver chalcogenides \cite{Rosenbaum1997,Zhang2011}, however, a linear magnetoresistance (LMR) is observed whose origin is still widely debated.
While it is frequently claimed to arise from a complicated band structure e.g. from a linear dispersion, more basic explanations \cite{Abrikosov1998} like sample inhomogeneities \cite{Parish2003,Narayanan2015,Hu2008}, density or sample thickness variations \cite{Bruls1981} may equally be responsible for the LMR. 
In order to distinguish between these different effects it would be desirable to explore the LMR phenomenon in a system with minimal disorder and a simple band structure.\\

In this Letter, we investigate an ultra-high mobility GaAs quantum well (QW) that can be described by an ideal free-electron-like model with a parabolic dispersion. We observe a strikingly large LMR up to 33\,T with a magnitude $(R(B)-R(0))/R(0)$ of order $10^5$\% onto which quantum oscillations become superimposed in the quantum Hall regime at low temperature. 
The free electron-like band structure, in combination with a nearly defect-free environment, excludes most of the possible exotic explanations for the appearance of a LMR. The quasi-two dimensionality of the transport additionally simplifies the situation such that sample thickness variations and surface/edge effects (e.g. as proposed in TIs) can be neglected or eliminated. Our results demonstrate that even in an ultra-clean free electron gas in two dimensions, a strikingly large LMR can be observed over a wide temperature range. Our subsequent analysis points to density fluctuations as the primary origin of the phenomenon.\\
Our sample is a 27\,nm-thick ultra-high mobility ($\bar{\mu}\approx 25\times 10^6$\,cm$^2$\,V$^{-1}$\,s$^{-1}$) $n$-type GaAs QW with dimensions of $4.5\times 4.5$\,mm$^2$. The sample was cleaved from a wafer that was grown by molecular beam epitaxy (MBE) without any rotation during the growth.
For the magnetotransport measurements, eight indium contacts were diffused in a van der Pauw geometry. Because of the low mean sheet density of $\bar{n}\approx 3 \times 10^{11}$\,cm$^{-2}$  only one single sub-band is occupied at low temperatures.  
We apply a small AC excitation current ($I\leq 100$\,nA) and measure the Hall and corresponding longitudinal signals of the sample in a standard six-point configuration with lock-in detection while sweeping the magnetic field perpendicular to the QW plane in a range from 0 to 33\,T.\\
\begin{figure}
 \includegraphics[width=\linewidth]{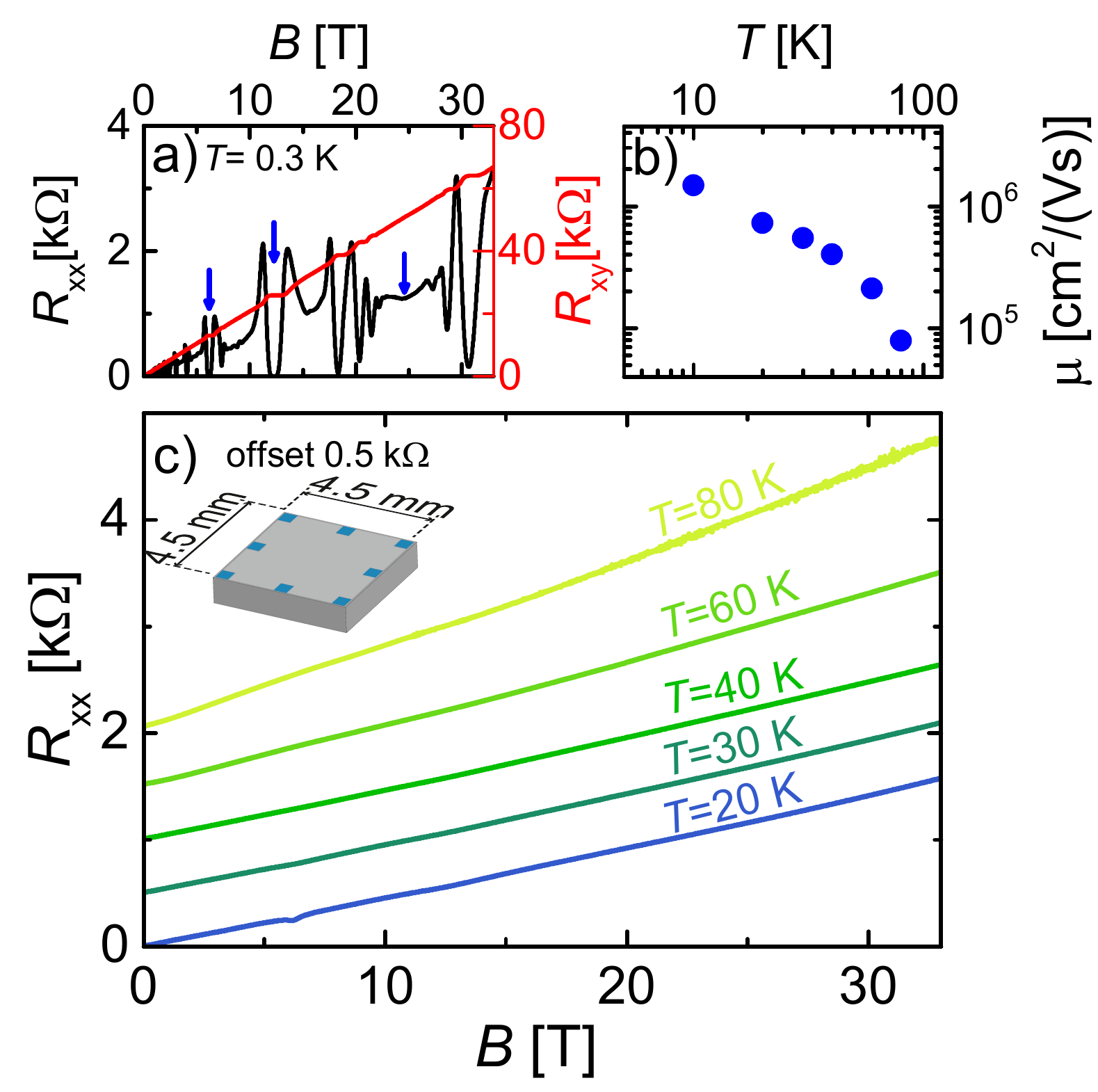}%
 \caption{\label{Fig1}  a) Longitudinal (black) and Hall resistance (red) at 0.3\,K. The blue arrows indicate from left to right the positions of the filling factors $\nu = 2,1$ and 1/2, respectively. b) Drude mobility as calculated from the sheet resistance at zero magnetic field. c) Longitudinal resistance curves in a temperature range from 20\,K to 80\,K. Different temperatures are indicated by different colours and are offset by a constant value of 500\,$\Omega$ for clarity. The inset shows a schematic of the sample dimensions}
 \end{figure}
In Fig.\ref{Fig1}~c) we present measurements of the longitudinal resistance $R_\textnormal{xx}$ from 20\,K to 80\,K. Even in the absence of quantum effects, a large LMR is observed extending down to almost zero magnetic field with a magnitude that exhibits a negligible temperature dependence.
The temperature independence of the LMR is remarkable given that the sample mobility varies by a factor of 10 over the same temperature range, as shown in Fig.~\ref{Fig1}~b). 
At low temperatures ($T=0.3$\,K), the usual Shubnikov-de Haas oscillations are observed in $R_\textnormal{xx}$ and Hall plateaus in $R_\textnormal{xy}$ associated with the integer and fractional quantum Hall regimes (see Fig.~\ref{Fig1}~a)).
Interestingly this vast spectrum of (fractional) quantum Hall states evolves out of a LMR background with a slope that is similar to the one seen at high temperatures.
This is illustrated in Fig.~\ref{Fig2}~a) where we show representative curves between $T=$ 0.3\,K and 20\,K. The 20\,K curve is reproduced here to demonstrate the similarities of the LMR at high and low temperatures. We emphasize that the very strong LMR that we observe over a wide temperature range is particularly striking given the extremely low defect concentration in our material system ($l_\textnormal{mfp}> 200$\,\textmu m).\\
A linear component in the longitudinal resistance at low temperatures is known to arise in 2DEGs from the so-called resistance rule that states that $R_\textnormal{xx}= \frac{\textnormal{d}R_\textnormal{xy}}{\textnormal{d}B} \times B \times \alpha = R_\textnormal{diff}$ \cite{Pan2005,Chang1985,Stormer1992,Rotger1989,Simon1994} where $\alpha$ is the constant of proportionality. In Fig.~\ref{Fig2}~a) we show that this rule is well obeyed in our sample with $\alpha \sim$ 0.026 $\pm 0.003$ across the entire temperature range. It is noted here from previous work that $\alpha$ has almost the same value for samples with mobilities and densities varying by two orders of magnitude \cite{Tieke1997}. At high temperatures, in the absence of Hall plateaus, $\frac{\textnormal{d}R_\textnormal{xy}}{\textnormal{d}B}$ becomes a constant and the empirical resistance rule describes the featureless LMR equally well. It is significant that this same resistance rule is valid over the entire temperature range with the same value of $\alpha$. 
Indeed, most classical transport theories \cite{Bruls1981,VanGelder1978} explain the LMR and the resistance rule as an admixture of a component of the Hall resistivity (which depends linearly on the magnetic field) to the longitudinal resistance caused by inhomogeneities such as density or sample thickness variations. Such variations give rise to a gradient of the (transverse) Hall voltage in the {\it longitudinal} direction that will naturally be picked up in measurements of $R_\textnormal{xx}$, therefore giving rise to a linear component in the magnetoresistance.
At 0.3\,K, we find that the location of the Hall plateaus and the minima in the longitudinal resistance do not coincide exactly with respect to the magnetic field position allowing us to estimate roughly a small density gradient of $\Delta n\approx 0.7 \times 10^{10}$\,cm$^{-2}$ across the sample amounting to a variation of $\sim$ 2.3\% along the length of the sample. Although this deviation is rather small, it can severely influence the measurements of $R_\textnormal{xx}$, especially in high mobility samples with a low mean sheet density like ours, where the longitudinal resistivity is small and the Hall voltage rises steeply with increasing $B$. This implies that $R_\textnormal{xx}$ is almost solely determined by the contribution from the Hall resistance $\Delta R_\textnormal{xy}$. In the simplest case, whereby we assume that only two different densities are present in the sample, we calculate $\Delta R_\textnormal{xy}=((1/n) - (1/(n+\Delta n)))\times \frac{B}{e}= 1578\,\Omega$ for $B=$ 33\,T, in good agreement with the measured value of $R_\textnormal{xx}$ of $1640\pm 85 \Omega$ (2.2\,K $\leq T \leq 40$\,K). \\
At even higher temperatures ($T > 60$\,K), the thermal energy is large enough to allow charge carriers to occupy the second sub-band of the quantum well which, according to self-consistent calculations, lies merely 70\,K above the Fermi energy. At this point the pure 2D-character of the system breaks down and the MR is governed by two sub-bands leading to deviations from the strict linear behavior, even though the density gradient remains (see Fig.~\ref{Fig1}~c)).

\begin{figure}
 \includegraphics[width=\linewidth]{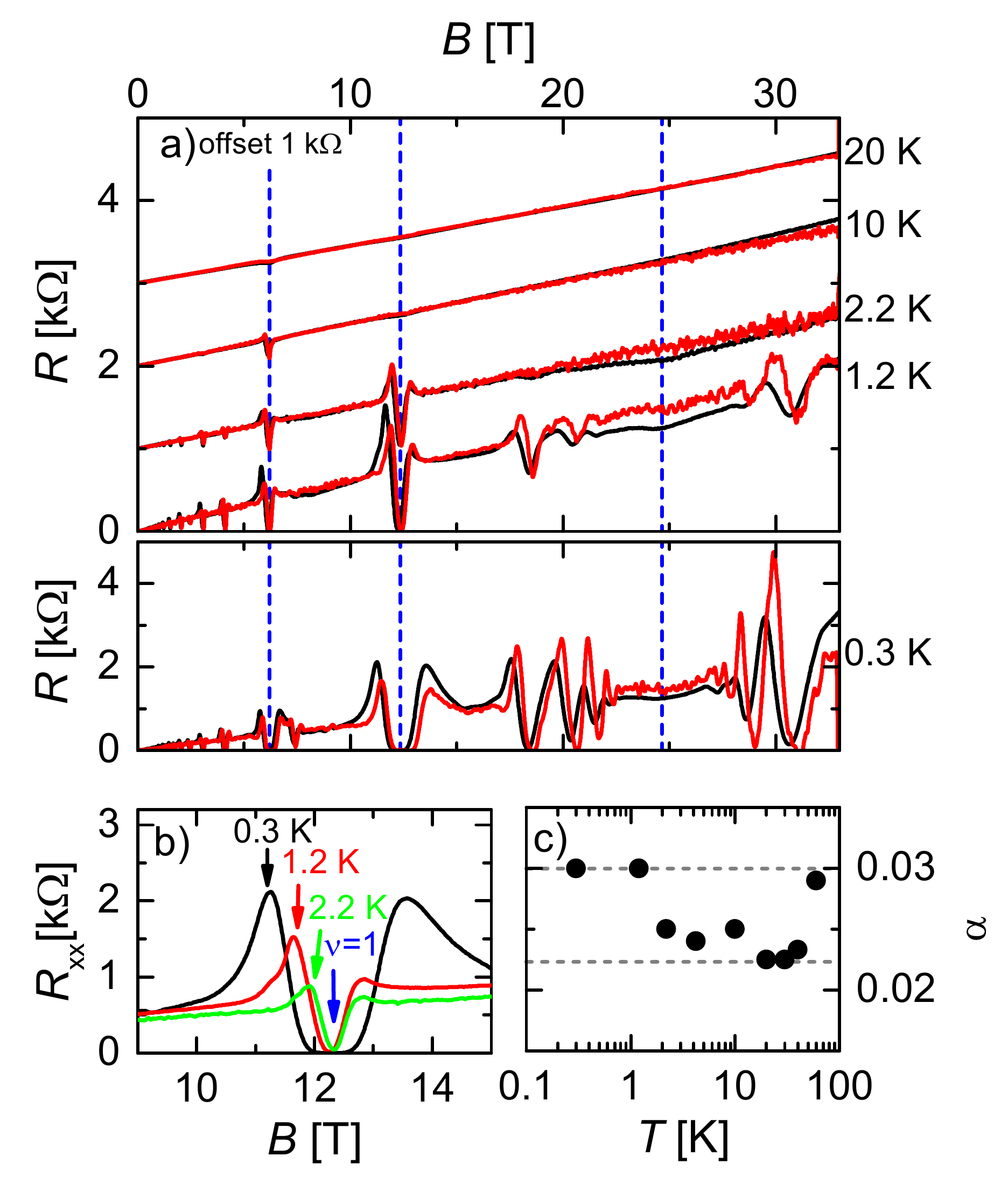}%
 \caption{\label{Fig2} a) Longitudinal resistance $R_\textnormal{xx}$ (black curve) and calculated resistance rule $R_\textnormal{diff}$ (red curve) for temperature of $T= 0.3,1.2,2.2,10$ and 20\,K. For clarity the curves have been offset with respect to each other and the (blue) dashed lines indicate from left to right the positions of the filling factors $\nu = 2,1$ and 1/2, respectively. b) Magnification of the region around filling factor $\nu =1$ for low temperatures. The position of $\nu =1$ is indicated by the blue arrow. c) Temperature dependence of the proportionality constant $\alpha$ as extracted from our data. The dashed lines show the spread of the data points.}
 \end{figure}
 
\begin{figure}
 \includegraphics[width=\linewidth]{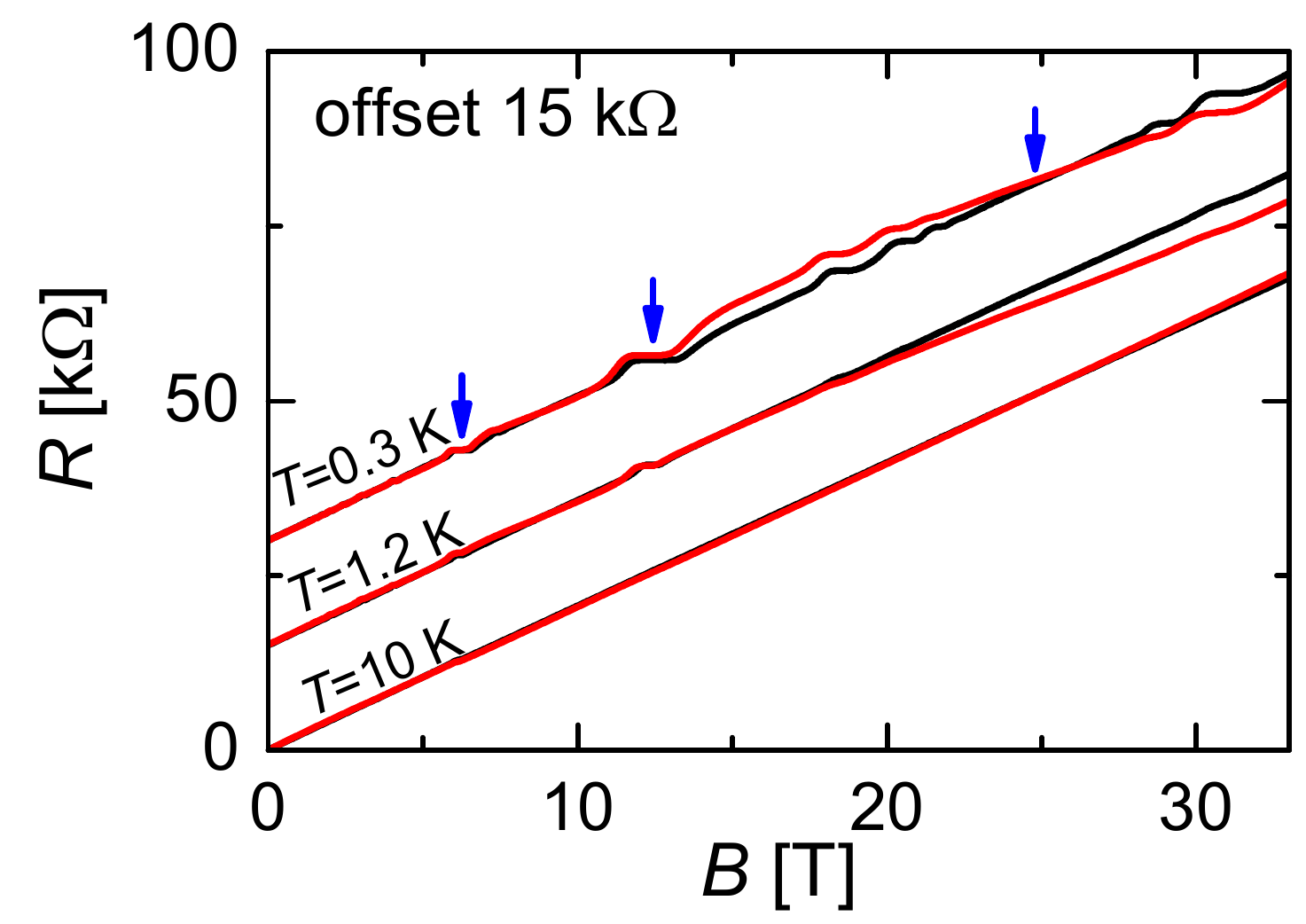}%
 \caption{\label{Fig3} Comparison of $R_\textnormal{xy}$ (black) and the inverted resistance rule $R_\textnormal{diff}^{-1}$ (red) for 0.3\,K,1.2\,K and 10\,K. For clarity the curves have been offset with respect to each other. The blue arrows indicate from left to right the positions of the filling factors $\nu = 2,1$ and 1/2, respectively.}
 \end{figure}

At low temperatures where the quantum oscillations are strong, we find that the resistance rule also provides a natural explanation for the maxima which appear before and after each minima. A magnification around $\nu= 1$ is representatively shown in Fig.~\ref{Fig2}~b). While the minima and zeroes in the resistance are commonly explained by the opening of an energy gap, the resistance maxima lack a detailed description. The derivative of the Hall resistance, however, which reproduces these features {\it including their temperature dependence} suggests that the presence of these maxima is closely linked to the density gradient across the sample.\\
Moreover, high magnetic fields enable us to investigate the region beyond $\nu= 1$ which to the best of our knowledge has not been studied with respect to the resistance rule thus far. Around the filling factor $\nu= 1/2$, deviations from the resistance rule are most obvious as  $R_\textnormal{diff}$ first slightly underestimates and then overestimates $R_\textnormal{xx}$ (see Fig.~\ref{Fig2}~a)). This difference disappears with increasing temperature and on approaching the classical transport regime where the quantum oscillations are suppressed. It is important to study such deviations from the resistance rule in order to understand its limitations and the underlying physics more deeply, especially as most previous studies mainly focused on the similarities between $R_\textnormal{xx}$ and $R_\textnormal{diff}$.\\
To proceed, we "invert" the resistance rule and calculate $R_\textnormal{xy}$ from $R_\textnormal{xx}$:
\begin{equation}
R_\textnormal{xy}= \int R_\textnormal{xx} B^{-1} \alpha ^{-1} \textnormal{d}B = R_\textnormal{diff}^{-1}
\label{invrrule}
\end{equation}
This approach has several advantages: Firstly, in contrast to the strength of quantum oscillations, the Hall resistance has well defined values in a magnetic field (including the high accuracy of the plateau values) thereby making it easier to identify deviations between $R_\textnormal{xy}$ and $R_\textnormal{diff}^{-1}$. 
Secondly, no derivatives are involved reducing the noise level of the calculated   $R_\textnormal{diff}^{-1}$ compared to the usual $R_\textnormal{diff}$.
It is of course necessary to add an integration constant $C$ which for our data set remains of order of 250 $\Omega$. 
The corresponding $\alpha (T)$ values match those determined from $R_{\textnormal{diff}}$. 
The comparison between $R_\textnormal{diff}^{-1}$ and $R_\textnormal{xy}$ is presented in Fig.~\ref{Fig3} where clear deviations are visible in the lowest Landau level at the lowest temperature. It is very plausible that these deviations are linked to the presence of composite fermions present at high magnetic fields and low temperatures. It is interesting that the deviations, of opposite sign at 0.3\,K, cross $R_\textnormal{xy}$ close to $\nu =1/2$.
We speculate that this crossing might reflect the opposite sign of the effective field $B_\textnormal{eff}$ of composite fermions. At high temperatures ($T\geq 10\,K$) where the fractional quantum Hall states are suppressed, both curves overlay each other almost perfectly.
Intriguingly, similar deviations from the resistance rule are apparent in Ref.\cite{Pan2005} around filling factor 3/2 which, however, are not addressed in this work.

In summary, we demonstrate the observation of a linear magnetoresistance in a high quality quasi-two-dimensional free electron gas with a parabolic dispersion. 
We show that this LMR is a generic phenomenon that is caused by an admixture of a component of the Hall resistivity to the longitudinal resistance and is strongly linked to density fluctuations inevitably present in most material systems and occurs even in ultra-high mobility 2DEGs with low carrier density.
Therefore, extreme care must be taken to attribute the observation of a LMR to more exotic effects or to complexities in the band structure. 
At low temperatures, quantum oscillations are superimposed on top of the LMR giving a structure that can be reproduced extremely well by the empirical resistance rule that mixes longitudinal and Hall components of the resistance tensor due to a small density gradient. One other key finding is that the resistance rule remains valid up to 60\,K where the thermal energy might have been expected to erase effects of the small density gradient in transport experiments.
Deviations from the resistance rule are observed at low temperatures and high magnetic fields ($\nu < 1 $) which might be related to new physics associated with the peculiar transport mechanism of composite fermions.

\bibliography{Literature}

\end{document}